\documentclass[envcountsame]{llncs}

\usepackage{amssymb}
\title{{\bf  Undecidable problems \\ about  timed automata} }

\author{Olivier Finkel\inst{}}
\institute{Equipe de Logique Math\'ematique \\
 U.F.R. de Math\'ematiques, Universit\'e Paris 7 \\ 2 Place Jussieu 75251 Paris
 cedex 05, France \\ \email{finkel@logique.jussieu.fr}.}

\date{}

\begin{document}

\spnewtheorem{Rem}[theorem]{Remark}{\bfseries}{\itshape}
\spnewtheorem{Exa}[theorem]{Example}{\bfseries}{\itshape}

\spnewtheorem{Pro}[theorem]{Proposition}{\bfseries}{\itshape}
\spnewtheorem{Lem}[theorem]{Lemma}{\bfseries}{\itshape}
\spnewtheorem{Cor}[theorem]{Corollary}{\bfseries}{\itshape}
\spnewtheorem{Deff}[theorem]{Definition}{\bfseries}{\itshape}

\newcommand{\fa}{\forall}
\newcommand{\Ga}{\Gamma}
\newcommand{\Gas}{\Gamma^\star}
\newcommand{\Gao}{\Gamma^\omega}

\newcommand{\Si}{\Sigma}
\newcommand{\Sis}{\Sigma^\star}
\newcommand{\Sio}{\Sigma^\omega}
\newcommand{\ra}{\rightarrow}
\newcommand{\hs}{\hspace{12mm}

\noi}
\newcommand{\lra}{\leftrightarrow}
\newcommand{\la}{language}
\newcommand{\ite}{\item}
\newcommand{\Lp}{L(\varphi)}
\newcommand{\abs}{\{a, b\}^\star}
\newcommand{\abcs}{\{a, b, c \}^\star}
\newcommand{\ol}{ $\omega$-language}
\newcommand{\orl}{ $\omega$-regular language}
\newcommand{\om}{\omega}
\newcommand{\nl}{\newline}
\newcommand{\noi}{\noindent}
\newcommand{\tla}{\twoheadleftarrow}
\newcommand{\de}{deterministic }
\newcommand{\proo}{\noi {\bf Proof.} }
\newcommand {\ep}{\hfill $\square$}

\maketitle

\begin{abstract}

\noi We solve some decision problems for timed automata which were 
raised by S. Tripakis in \cite{tri} and by E. Asarin in \cite{asa}. In particular,  
we show that one cannot decide whether a given timed automaton is determinizable 
or whether the complement of a timed regular language is timed regular. 
We show that the problem of the minimization of the number of clocks of a timed automaton 
is undecidable. 
It is also undecidable whether the shuffle of two  timed regular languages is  timed regular. 
We show that in the case of timed B\"uchi automata accepting  infinite timed words some of these 
problems are   $\Pi_1^1$-hard,  hence highly undecidable (located beyond the arithmetical hierarchy).\footnote{Part of the results 
stated in this paper were presented very  recently in the Bulletin of the EATCS \cite{beatcs1,beatcs2}. }  

\end{abstract}

\noi {\small {\bf Keywords:} Timed automata; timed B\"uchi automata; timed regular ($\om$)-languages; 
decision problems; universality problem; determinizability; 
complementability; shuffle operation; minimization of the number of clocks.} 

\section{Introduction}

R. Alur and D.  Dill    introduced in  \cite{ad}  the notion of timed automata reading 
timed words. Since then the theory of timed automata has been much studied and used for specification and verification of timed systems. 
\nl In a recent paper, E. Asarin   raised a series of questions about the theoretical foundations of timed automata and timed languages 
which were still open  and  wrote: ``I believe that getting answers to them would substantially improve our understanding 
of the area" of timed systems,  \cite{asa}. 
\nl Some of  these questions  concern  decision problems ``\`{a} la \cite{tri}".  For instance : ``Is it possible, given a timed automaton 
$\mathcal{A}$, to decide whether it is equivalent to a deterministic one ?". 
\nl S. tripakis showed  in \cite{tri} that there is no algorithm which, given a timed automaton 
$\mathcal{A}$,  decides whether it is equivalent to a deterministic one,  {\bf and} if this is the case 
gives an equivalent deterministic  automaton $\mathcal{B}$. But the above question of the decidability of the determinizability 
alone (where we do not require the construction of the witness $\mathcal{B}$) was still open. 
\nl  We give in this paper the answer to this question and to 
several other ones  of  \cite{tri,asa}.  In particular,  
we show that one cannot decide whether a given timed automaton is determinizable 
or whether the complement of a timed regular language is timed regular. 
We study also the corresponding problems but with ``bounded resources" stated in \cite{tri}. 
\nl  For that purpose we use a  method which is very similar to that one used in \cite{rel-dec} to prove undecidability 
results about infinitary rational relations, reducing  the universality problem, which is undecidable,  to some other decision problems. 
\nl  We study also the problem of  the minimization of the number of clocks of a timed automaton, showing that 
one cannot decide, for a given timed automaton $\mathcal{A}$ with $n$ clocks, $n\geq 2$, whether there is an equivalent 
timed automaton $\mathcal{B}$ with at most $n-1$ clocks. 
\nl The question of the closure 
of the class of timed regular languages under shuffle was  also raised by E. Asarin in \cite{asa}.  
C. Dima proved in \cite{dim} that timed regular expressions with shuffle characterize timed languages accepted by stopwatch automata. This implies that 
the class of timed regular languages is not closed under shuffle. We proved this result independently in \cite{beatcs2}. We  recall the proof here, 
giving a simple example of two timed regular languages whose shuffle 
is not timed regular. Next we use this example to 
prove that one can not decide whether the shuffle of two given timed regular languages is 
 timed regular. 
\nl We extend also the previous undecidability results  to the case of timed B\"uchi automata accepting infinite timed 
words. In this case many problems are  $\Pi_1^1$-hard,  hence highly undecidable (located beyond the arithmetical hierarchy), 
because   the universality 
problem for timed B\"uchi automata, which is itself  $\Pi_1^1$-hard, \cite{ad}, can be reduced to these other decision problems.
\nl We mention that part of the results stated in this paper were presented very  recently in the Bulletin of the EATCS \cite{beatcs1,beatcs2}. 

\hs The paper is organized as follows. We recall usual notations in Section 2. 
 The undecidability of determinizability or 
regular  complementability for timed regular languages is proved in Section 3. 
The problem of minimization of the number of clocks is studied in Section 4. 
Results about the shuffle operation are stated in Section 5. Finally we extend in Section 6 some 
undecidability results  to the case of timed B\"uchi automata.

\section{Notations}

We assume the reader to be familiar with the basic  theory of timed languages and 
timed automata (TA) \cite{ad}. 
\nl   The set of positive reals will be denoted $\mathcal{R}$. 
A (finite length) timed word over  a finite alphabet $\Si$ is of the form 
$t_1.a_1.t_2.a_2 \ldots t_n.a_n$, where, for all integers $i\in [1, n]$, 
 $t_i \in \mathcal{R}$ and $a_i \in \Si$. 
It may be seen as a {\it time-event sequence}, where the $t_i \in  \mathcal{R}$ represent 
time lapses between events and the letters $a_i \in \Si$ represent events. 
The set of all (finite length) timed words 
over a finite alphabet $\Si$ is the set $(\mathcal{R} \times \Si)^\star$. A timed language 
is a subset of $(\mathcal{R} \times \Si)^\star$.
The complement ( in $(\mathcal{R} \times \Si)^\star$ ) of a timed language 
$L \subseteq (\mathcal{R} \times \Si)^\star$ is $(\mathcal{R} \times \Si)^\star - L$ denoted 
$L^c$. 
\nl We consider a basic model of timed automaton, as introduced in \cite{ad}. 
A timed automaton $\mathcal{A}$ 
has a finite set of states and a finite set of transitions. Each 
transition is labelled with a letter of a finite input alphabet $\Si$. We assume 
that each transition of $\mathcal{A}$ has a set 
of clocks to reset to zero and only {\it diagonal-free} clock guard \cite{ad}. 
\nl A timed automaton $\mathcal{A}$ is said to be 
deterministic iff it satisfies the two following requirements: 
\nl (a) $\mathcal{A}$ has only one start state,  and 
\nl (b)  if there are multiple transitions starting at the same state 
with the same label, then their clock constraints are mutually exclusive. 
\nl Then 
a deterministic timed automaton $\mathcal{A}$ has at most one run on a given timed word \cite{ad}. 
\nl As usual, we denote by $L(\mathcal{A})$ the timed language accepted (by final states) by the 
timed automaton  $\mathcal{A}$. A timed language $L \subseteq (\mathcal{R} \times \Si)^\star$ 
is said to be timed regular iff there is a timed automaton  $\mathcal{A}$ such that $L=L(\mathcal{A})$. 

\hs An infinite timed word over  a finite alphabet $\Si$ is of the form 
$t_1.a_1.t_2.a_2.t_3.a_3 \ldots$, where, for all integers $i\geq 1$, 
 $t_i \in \mathcal{R}$ and $a_i \in \Si$. 
It may be seen as an infinite  {\it time-event sequence}.  
The set of all   infinite timed words 
over $\Si$ is the set $(\mathcal{R} \times \Si)^\om$. A timed $\om$-language 
is a subset of $(\mathcal{R} \times \Si)^\om$.
The complement ( in $(\mathcal{R} \times \Si)^\om$ ) of a timed $\om$-language 
$L \subseteq (\mathcal{R} \times \Si)^\om$ is $(\mathcal{R} \times \Si)^\om - L$ denoted 
$L^c$. 
\nl We consider a basic model of  timed  B\"uchi  automaton, (TBA), as introduced in \cite{ad}. 
We assume, as in the case of TA accepting finite timed words, 
that each transition of $\mathcal{A}$ has a set 
of clocks to reset to zero and only {\it diagonal-free} clock guard \cite{ad}. 
 The timed $\om$-language accepted  by the 
timed B\"uchi  automaton  $\mathcal{A}$  is  denoted $L_\om(\mathcal{A})$. 
A timed language $L \subseteq (\mathcal{R} \times \Si)^\om$ 
is said to be timed $\om$-regular iff there is a timed B\"uchi  automaton  $\mathcal{A}$ such that $L=L_\om(\mathcal{A})$. 

\section{Complementability and determinizability}

 We first state   the undecidability of determinizability or 
regular  complementability for timed regular languages. 

\begin{theorem}
It is undecidable to determine, for a given TA $\mathcal{A}$, whether 
\begin{enumerate}
\ite $L(\mathcal{A})$ is accepted by a deterministic TA.
\ite $L(\mathcal{A})^c$ is accepted by a TA. 
\end{enumerate}
\end{theorem}

\proo It is well known that the class of timed regular languages is not closed under 
complementation. Let $\Si$ be a finite alphabet and let $a\in \Si$. 
Let $A$ be the set of timed words of the form 
$t_1.a.t_2.a \ldots t_n.a$, where, for all integers $i\in [1, n]$, 
$t_i \in \mathcal{R}$ and there is a pair of integers $(i, j)$ such that $i, j \in [1, n]$,  
$i<j$, and $t_{i+1}+t_{i+2}+ \ldots +t_j = 1$. The timed language $A$ is formed by timed 
words containing only letters $a$ and such that there is a pair of  $a$'s which are separated 
by a time distance $1$. The timed language $A$ is regular but its  complement 
 can not be accepted by any timed automaton because such an 
 automaton should have an unbounded number of clocks to check that no pair of  
$a$'s  is separated by a time distance $1$, \cite{ad}. 

\hs We shall use the undecidability of the universality problem for timed regular languages: 
one cannot decide, for a given timed automaton  $\mathcal{A}$ with input alphabet $\Si$, 
whether $L(\mathcal{A})=(\mathcal{R} \times \Si)^\star$,  \cite{ad}. 

\hs Let $c$ be an additional  letter not in $\Si$. 
 For a given timed regular language $L \subseteq (\mathcal{R} \times \Si)^\star$, we are going to  construct another 
timed language $\mathcal{L}$ over the alphabet $\Ga=\Si\cup\{c\}$ defined as the union of 
the following three languages. 

\begin{itemize}
\ite $\mathcal{L}_1 = L.(\mathcal{R} \times \{c\}).(\mathcal{R} \times \Si)^\star$
\ite  $\mathcal{L}_2$ is the set of timed words over $\Ga$ having  no $c$'s or having at 
least two $c$'s.  
\ite $\mathcal{L}_3 = (\mathcal{R} \times \Si)^\star.(\mathcal{R} \times \{c\}).A$, where $A$ is the above defined timed regular 
language over the alphabet $\Si$. 
\end{itemize}  

\hs The timed  language  $\mathcal{L}$ is regular because $L$ and $A$ are regular timed languages. 
 There are  now two cases. 
\begin{enumerate}

\ite[(1)] {\bf First case.}  $L = (\mathcal{R} \times \Si)^\star$. Then $\mathcal{L}= (\mathcal{R} \times (\Si\cup\{c\}))^\star$. 
Therefore $\mathcal{L}$ has the minimum possible complexity. $\mathcal{L}$ is of course accepted by a deterministic timed automaton 
(without any clock). 
Moreover its complement $\mathcal{L}^c$ is empty thus it is also accepted by a deterministic timed automaton 
(without any clock). 

\ite[(2)] {\bf Second case.} $L$ is strictly included into  $(\mathcal{R} \times \Si)^\star$. Then there is a timed word 
$u=t_1.a_1.t_2.a_2  \ldots  t_n.a_n \in (\mathcal{R} \times \Si)^\star$ which does not belong to $L$. Consider now a timed word 
$x \in (\mathcal{R} \times \Si)^\star$.  It holds that $u.1.c.x \in \mathcal{L}$ iff $x\in A$. 
 Then we have also :  $u.1.c.x \in \mathcal{L}^c$ iff $x\in A^c$. 
\nl We are going to show that $\mathcal{L}^c$ is not timed regular. Assume on the contrary that there is a timed automaton 
$\mathcal{A}$ such that $\mathcal{L}^c=L(\mathcal{A})$.  There are only finitely many possible global states (including the clock values)  of 
$\mathcal{A}$ after the reading of the initial segment $u.1.c$. It is clearly not possible that the timed automaton $\mathcal{A}$, from these global states, 
accept all timed words in $A^c$ and only these ones, for the same reasons which imply that $A^c$ is not timed regular. 
Thus  $\mathcal{L}^c$ is not timed regular. This implies that  $\mathcal{L}$ is not accepted by any deterministic timed automaton because the 
class of deterministic regular timed languages is closed under complement. 
\end{enumerate}

\noi In the first case  $\mathcal{L}$ is accepted by a deterministic timed automaton and $\mathcal{L}^c$ is timed regular. 
In the second case $\mathcal{L}$ is not accepted by any deterministic timed automaton and $\mathcal{L}^c$ is not  timed regular. 
But one cannot decide which case holds because of the undecidability of the universality problem for timed regular languages. 
\ep

\hs  Below $TA(n, K) $ denotes the class  of timed automata having 
at most $n$ clocks and where constants are at most $K$.
In \cite{tri}, Tripakis stated the following problems which are similar to the above ones but with 
``bounded resources".

\hs Problem 10  of \cite{tri}. Given a TA $\mathcal{A}$ and non-negative integers $n, K$, does there exist a TA
$\mathcal{B} \in  TA(n, K)$ such that $L(\mathcal{B})^c = L(\mathcal{A})$ ? If so, construct such a $\mathcal{B}$.

\hs Problem 11  of \cite{tri}. Given a TA $\mathcal{A}$ and non-negative integers $n, K$, does there exist a
deterministic TA $\mathcal{B} \in TA(n, K)$  such that $L(\mathcal{B}) = L(\mathcal{A})$ ? 
If so, construct such a
$\mathcal{B}$.

\hs Tripakis showed that these problems are not algorithmically solvable. He asked also 
 whether these bounded-resource versions of previous problems remain undecidable if we 
do not require the construction of the witness $\mathcal{B}$, i.e. if we 
omit the sentence 
``If so construct such a $\mathcal{B}$" in the statement of Problems 10 and 11. 
\nl It is easy to see, from the proof of preceding Theorem,  that this is actually the case because 
we have seen that, in the first case, $\mathcal{L}$ and $\mathcal{L}^c$ are accepted by deterministic 
timed automata {\it without any clock}.  

\section{Minimization of the number of clocks}

\noi The following problem was  shown to be undecidable by S. Tripakis in \cite{tri}.

\hs Problem 5 of \cite{tri}. Given a TA $\mathcal{A}$ with $n$ clocks, does there exists a TA $\mathcal{B}$ with $n-1$
clocks, such that $L(\mathcal{B}) = L(\mathcal{A})$ ? If so, construct such a $\mathcal{B}$.

\hs The corresponding decision problem, where we require only a Yes / No answer but no witness in the case 
of a positive answer, was left open in \cite{tri}. 
\nl Using a very similar reasoning as in the preceding section, we can prove that this problem is  also 
undecidable. 

\begin{theorem}\label{min} Let $n \geq 2$ be a positive integer. 
It is undecidable to determine, for a given TA $\mathcal{A}$ with $n$ clocks, whether  there exists a TA $\mathcal{B}$ with $n-1$
clocks, such that $L(\mathcal{B}) = L(\mathcal{A})$.
\end{theorem}

\proo  Let $\Si$ be a finite alphabet and let $a\in \Si$. Let $n \geq 2$ be a positive integer, and 
 $A_n$ be the set of timed words of the form 
$t_1.a.t_2.a \ldots t_k.a$, where, for all integers $i\in [1, k]$, 
$t_i \in \mathcal{R}$ and there are $n$  pairs of integers $(i, j)$ such that $i, j \in [1, k]$,  
$i<j$, and $t_{i+1}+t_{i+2}+ \ldots +t_j = 1$. The timed language $A_n$ is formed by timed 
words containing only letters $a$ and such that there are $n$  pairs of  $a$'s which are separated 
by a time distance $1$. 
$A_n$ is a timed regular language but it can not be accepted by any timed automaton with less than $n$ clocks, see 
\cite{hkw}. 

\hs Let $c$ be an additional  letter not in $\Si$. 
 For a given timed regular language $L \subseteq (\mathcal{R} \times \Si)^\star$ 
accepted by a TA with at most $n$ clocks, we construct another 
timed language $\mathcal{V}_n$ over the alphabet $\Ga=\Si\cup\{c\}$ defined as the union of 
the following three languages. 

\begin{itemize}
\ite $\mathcal{V}_{n,1} = L.(\mathcal{R} \times \{c\}).(\mathcal{R} \times \Si)^\star$
\ite  $\mathcal{V}_{n,2}$ is the set of timed words over $\Ga$ having  no $c$'s or having at 
least two $c$'s. 
\ite $\mathcal{V}_{n,3} = (\mathcal{R} \times \Si)^\star.(\mathcal{R} \times \{c\}).A_n$. 
\end{itemize}  

\hs The timed  language  $\mathcal{V}_n$ is regular because $L$ and $A_n$ are regular timed languages. 
Moreover it is easy to see that $\mathcal{V}_n$ is accepted by a TA with at most $n$ clocks, because $L$ and $A_n$ are 
accepted by timed automata with at most $n$ clocks.
 There are  now two cases. 
\begin{enumerate}

\ite[(1)] {\bf First case.}  $L = (\mathcal{R} \times \Si)^\star$. Then $\mathcal{V}_n= (\mathcal{R} \times (\Si\cup\{c\}))^\star$,  
thus  $\mathcal{V}_n$ is accepted by a (deterministic) timed automaton 
{\it without any clock}. 

\ite[(2)] {\bf Second case.} $L$ is strictly included into  $(\mathcal{R} \times \Si)^\star$. Then there is a timed word 
$u=t_1.a_1.t_2.a_2  \ldots  t_k.a_k \in (\mathcal{R} \times \Si)^\star$ which does not belong to $L$. Consider now a timed word 
$x \in (\mathcal{R} \times \Si)^\star$.  It holds that $u.1.c.x \in \mathcal{V}_n$ iff $x\in A_n$. 
\nl Towards a contradiction, assume that  $\mathcal{V}_n$ is accepted  by a timed automaton $\mathcal{B}$ with at most $n-1$ clocks. 
 There are only finitely many possible global states (including the clock values)  of 
$\mathcal{B}$ after the reading of the initial segment $u.1.c$. It is clearly not possible that the timed automaton $\mathcal{B}$, from these global states, 
accept all timed words in $A_n$ and only these ones, because it has less than $n$ clocks.  
\end{enumerate}

\noi But one cannot decide which case holds because of the undecidability of the universality problem for timed regular languages 
accepted by timed automata with $n$ clocks, where $n\geq 2$. 
\ep 

\begin{Rem}
For timed automata with only one clock, the inclusion problem, hence also the universality problem, have recently been shown to 
be decidable by  J.  Ouaknine and J. Worrell   \cite{ow}. 
Then  the above method can not be applied.   It is easy to see that it is decidable whether a timed regular language 
accepted by a timed automaton with only one clock is also accepted by a timed automaton without any clock. 
\end{Rem}

\section{Shuffle operation}

\noi  It is well known that the class of timed regular languages is closed under union, intersection, but not under 
complementation. 
Another usual operation is the shuffle operation. Recall that the shuffle $x \Join y$ of two elements 
$x$ and $y$ of a monoid $M$ is the set of all products of the form 
$x_1\cdot y_1\cdot x_2\cdot y_2 \cdots x_n\cdot y_n$ where $x=x_1\cdot x_2 \cdots x_n$ and $y=y_1\cdot y_2 \cdots y_n$. 
\nl This operation can naturally be extended to subsets of $M$ by setting,  for $R_1, R_2 \subseteq M$, 
$R_1 \Join R_2 = \{ x \Join y \mid x \in R_1 \mbox{ and } y \in R_2 \}$.  
\nl We know that the class of regular (untimed) languages is  closed under shuffle.  The question of the closure 
of the class of timed regular languages under shuffle was  raised by E. Asarin in \cite{asa}.  
C. Dima proved in \cite{dim} that timed regular expressions with shuffle characterize timed languages accepted by stopwatch automata. This implies that 
the class of timed regular languages is not closed under shuffle. We proved this result independently in \cite{beatcs2}. 
\nl We are going to reprove this  here, 
giving a simple example of two timed regular languages whose shuffle 
is not timed regular. Next we shall use this example to 
prove that one cannot decide whether the shuffle of two given timed regular languages is 
 timed regular. 

\begin{theorem}
The shuffle of timed regular languages is not always timed regular. 
\end{theorem}

\proo  Let $a, b$ be two different letters and $\Si=\{a, b\}$. 
\nl Let $R_1$ be the language of timed words over $\Si$ of the form 
$$t_1\cdot a\cdot 1\cdot a\cdot t_2\cdot a$$
\noi for  some positive reals $t_1$ and $t_2$ such that $t_1+1+t_2=2$, i.e. $t_1+t_2=1$. 
\nl It is clear that $R_1$ is a timed regular language of finite timed words. 

\hs {\bf Remark.}
As remarked in \cite[page 217]{ad}, 
a timed automaton can compare delays with constants, but it cannot remember delays. 
If we would like a timed automaton 
to be able to compare delays, we should add clock constraints of the form $x+y \leq x'+y'$ for some clock values $x, y, x', y'$. 
But this would greatly increase the expressive power of automata: the languages accepted by such automata are not always timed regular, 
and if we allow the addition primitive in the syntax of clock constraints, 
then the emptiness problem for timed automata would be undecidable \cite[page 217]{ad}. 

\hs Notice that the above  language $R_1$ is timed regular because  a timed automaton 
$\mathcal{B}$ reading a word of the form $t_1\cdot a\cdot 1\cdot a\cdot t_2\cdot a$,  for some positive reals $t_1$ and $t_2$,  
can compare the delays $t_1$ and $t_2$ in order to check that $t_1+t_2=1$. This is due to the fact that the delay between the 
two first occurrences of the event $a$ is {\it constant } equal to 1. 
\nl Using the shuffle operation we shall construct a language  $R_1 \Join  R_2$,  for a regular timed language $R_2$. 
Informally speaking, this will  ``insert a variable delay" between the  two first occurrences of the event $a$ 
and the resulting  language $R_1 \Join  R_2$ will not be timed regular. 
\nl We now  give the details of this construction. 

\hs Let $R_2$ be the  language of timed words over $\Si$ of the form 
$$1\cdot b\cdot s\cdot b$$
\noi for some positive real $s$.
\nl The language $R_2$ is of course also a timed regular language. 

\hs We are going to prove that $R_1 \Join  R_2$ is not timed regular. 

\hs Towards a contradiction, assume that $R_1  \Join R_2$ is timed regular. 
Let $R_3$ be the set of timed words over $\Si$ of the form 
$$t_1\cdot a\cdot 1\cdot b\cdot s\cdot b\cdot 1\cdot a\cdot t_2\cdot a$$
\noi for some positive reals $t_1, s, t_2$. It is clear that $R_3$ is timed regular. 
On the other hand  the class of timed regular languages is closed under intersection thus 
the timed language  $(R_1  \Join R_2)  \cap R_3$ would be also timed regular. 
But this language is simply the set of timed words of the form $t_1\cdot a\cdot 1\cdot b\cdot s\cdot b\cdot 1\cdot a\cdot t_2\cdot a$, for 
some positive reals $t_1, s, t_2$ such that $t_1 + t_2 = 1$. 

\hs Assume that this timed language is accepted by a timed automaton $\mathcal{A}$. 

\hs Consider now  the reading by   $\mathcal{A}$  of a word of the form 
$t_1\cdot a\cdot 1\cdot b\cdot s\cdot b\cdot 1\cdot a\cdot t_2\cdot a$, for some positive reals $t_1, s, t_2$. 
\nl After reading  the initial segment $t_1\cdot a\cdot 1\cdot b\cdot s\cdot b\cdot 1\cdot a$  
the value of any clock of 
$\mathcal{A}$ can only be $t_1+s+2$, $2+s$, $1+s$, or $1$. 
\nl If the clock value of a clock $\mathcal{C}$ has been at some time reset to zero, its value 
may be $2+s$, $1+s$, or $1$. So the value $t_1$ is not stored in the clock value and this clock 
can not be used to compare $t_1$ and $t_2$ in order to check that $t_1+t_2 = 1$. 
\nl On the other hand if the clock value of a clock $\mathcal{C}$ has not been at some time reset to zero,  then, 
after  reading $t_1\cdot a\cdot 1\cdot b\cdot s\cdot b\cdot 1\cdot a$,   
its value will be $t_1+s+2$ .  This must hold for uncountably many 
values of the real $s$,  and again the value $t_1+s+2$ can not be used to accept, from the global 
state of $\mathcal{A}$ after  reading  the initial segment $t_1\cdot a\cdot 1\cdot b\cdot s\cdot b\cdot 1\cdot a$, only the word $t_2\cdot a$ for 
$ t_2 = 1-t_1 $. 

\hs This implies that  $(R_1  \Join R_2)  \cap R_3$ hence also $(R_1  \Join R_2)$ are not timed regular. 
\ep 

\hs We can now state the following result: 

\begin{theorem}\label{shuffle}
It is undecidable to determine whether the shuffle of two given timed regular languages is timed regular. 
\end{theorem}

\proo
\hs We shall use again the undecidability of the universality problem for timed regular languages: 
one cannot decide, for a given timed automaton  $\mathcal{A}$ with input alphabet $\Si$, 
whether $L(\mathcal{A})=(\mathcal{R} \times \Si)^\star$. 

\hs Let $\Si=\{a, b\}$, and $c$ be an additional  letter not in $\Si$. 
 For a given timed regular language $L \subseteq (\mathcal{R} \times \Si)^\star$, we are going firstly to  construct another 
timed language $\mathcal{L}$ 
over the alphabet $\Ga=\Si\cup\{c\}$. 

\hs The language  $\mathcal{L}$ is defined as the union of 
the following three languages. 

\begin{itemize}
\ite $\mathcal{L}_1 = L.(\mathcal{R} \times \{c\}).(\mathcal{R} \times \Si)^\star$
\ite  $\mathcal{L}_2$ is the set of timed words over $\Ga$ having  no $c$'s or having at 
least two $c$'s. 
\ite $\mathcal{L}_3 = (\mathcal{R} \times \Si)^\star.1.c.R_1$, 
where $R_1$ is the above defined timed regular 
language over the alphabet $\Si$. 
\end{itemize}  

\hs The timed  language  $\mathcal{L}$ is regular because $L$ and $R_1$ are regular timed languages. 
\nl Consider now the language $\mathcal{L}\Join R_2$, where  $R_2$ 
 is the above defined regular timed language. 
 
\hs  There are  now two cases. 
\begin{enumerate}

\ite[(1)] {\bf First case.}  $L = (\mathcal{R} \times \Si)^\star$. Then $\mathcal{L}= (\mathcal{R} \times (\Si\cup\{c\}))^\star$ and 
 $\mathcal{L}\Join R_2 = (\mathcal{R} \times (\Si\cup\{c\}))^\star$.  Thus 
 $\mathcal{L}\Join  R_2 $ is  timed regular. 

\ite[(2)] {\bf Second case.} $L$ is strictly included into  $(\mathcal{R} \times \Si)^\star$. 
\nl Towards a contradiction, assume that  $\mathcal{L}\Join  R_2 $ is  timed regular. 
Then the timed language $\mathcal{L}_4 = ( \mathcal{L}\Join  R_2 ) \cap [  (\mathcal{R} \times \Si)^\star.1.c.R_3 ]$, where $R_3$ is the 
above defined  timed regular language,  would be also timed regular because it would be the intersection of two timed regular languages.
\nl On the other hand  $L$ is strictly included into  $(\mathcal{R} \times \Si)^\star$ thus  there is a timed word 
$u=t_1.a_1.t_2.a_2  \ldots  t_n.a_n \in (\mathcal{R} \times \Si)^\star$ which does not belong to $L$. 
\nl Consider now a timed word 
$x \in (\mathcal{R} \times \Si)^\star$.  It holds that $u.1.c.x \in \mathcal{L}_4$ iff  $x\in ( R_1 \Join R_2 ) \cap R_3$.  
\nl We are going to show now  that $\mathcal{L}_4$ is not timed regular. Assume on the contrary that there is a timed automaton 
$\mathcal{A}$ such that $\mathcal{L}_4=L(\mathcal{A})$. 
 There are only finitely many possible global states (including the clock values)  of 
$\mathcal{A}$ after the reading of the initial segment $u.1.c$. It is clearly not possible that the timed automaton $\mathcal{A}$, from these global states, 
accept all timed words in   $ ( R_1 \Join R_2 ) \cap R_3$   and only these ones, 
for the same reasons which imply that $( R_1 \Join R_2 ) \cap R_3$ is not timed regular. 
Thus  $\mathcal{L}_4$ is not timed regular and this implies that $\mathcal{L}\Join R_2$ is not timed regular. 
\end{enumerate}

\noi In the first case  $\mathcal{L}\Join  R_2$ is  timed regular. 
In the second case $\mathcal{L}\Join R_2$ is not timed regular. 
But one cannot decide which case holds because of the undecidability of the universality problem for timed regular languages. 
\ep 

\hs We can also study the corresponding problems with ``bounded resources": 

\hs Problem 1. Given two timed automata  $\mathcal{A}$ and $\mathcal{B}$ and non-negative integers $n, K$, does there exist a TA
$\mathcal{C} \in  TA(n, K)$ such that $L(\mathcal{C}) = L(\mathcal{A}) \Join L(\mathcal{B}) $ ? 

\hs Problem 2. Given two timed automata  $\mathcal{A}$ and $\mathcal{B}$ and an  integer $n \geq 1$, does there exist a TA
$\mathcal{C}$ with less than $n$ clocks such that $L(\mathcal{C}) = L(\mathcal{A}) \Join L(\mathcal{B}) $ ? 

\hs Problem 3. Given two timed automata  $\mathcal{A}$ and $\mathcal{B}$, does there exist a deterministic TA
$\mathcal{C}$  such that $L(\mathcal{C}) = L(\mathcal{A}) \Join L(\mathcal{B}) $ ? 

\hs From the proof of above Theorem \ref{shuffle}, it is easy to see that these problems are also undecidable. Indeed in the first case 
$\mathcal{L}\Join  R_2$ was accepted by a deterministic timed automaton without any clocks. And in the second case 
$\mathcal{L}\Join  R_2$  was not accepted by any timed automaton. 

\hs E. Asarin, P. Carpi, and O. Maler have proved in \cite{acm} that  the formalism of timed regular expressions (with intersection and renaming) 
has the same expressive power than  timed  automata.  C. Dima  proved in \cite{dim} that timed regular expressions with shuffle 
characterize timed languages accepted by stopwatch automata. We refer the reader to  \cite{dim} for the definition of  stopwatch automata. 
\nl Dima showed that, from two timed automata  $\mathcal{A}$ and $\mathcal{B}$, one can construct a stopwatch automaton $\mathcal{C}$ such that 
$L(\mathcal{C}) = L(\mathcal{A}) \Join L(\mathcal{B}) $. 
Thus we can infer  the following corollaries from the above results. 
 \nl Notice that in \cite{acm,dim} the authors consider automata with epsilon-transitions while in this paper we have only considered timed automata 
without  epsilon-transitions, although we think that many results could be extended to the case of automata with epsilon-transitions. So in the statement of the following 
corollaries we consider   stopwatch automata with  epsilon-transitions but only timed automata without  epsilon-transitions.

\begin{Cor}
One cannot decide, for a given stopwatch automaton $\mathcal{A}$,  whether there exists a timed automaton $\mathcal{B}$ 
(respectively, a deterministic timed automaton $\mathcal{B}$)  such that 
$L(\mathcal{A}) = L(\mathcal{B})$. 
\end{Cor}

\begin{Cor}
One cannot decide, for a given stopwatch automaton $\mathcal{A}$ and  non-negative integers $n, K$,  whether
there exists a timed automaton $\mathcal{B} \in TA(n, K)$  such that 
$L(\mathcal{A}) = L(\mathcal{B})$. 
\end{Cor}

\begin{Cor}
One cannot decide, for a given stopwatch automaton $\mathcal{A}$ and an  integer $n \geq 1$,  whether
there exists a timed automaton $\mathcal{B}$  with less than $n$ clocks such that 
$L(\mathcal{A}) = L(\mathcal{B})$. 
\end{Cor}

\section{Timed B\"uchi automata}

The previous undecidability results can be extended to the case of timed B\"uchi automata accepting infinite timed 
words. Moreover in this case many problems are highly undecidable ($\Pi_1^1$-hard) because  the universality 
problem for timed B\"uchi automata, which is itself $\Pi_1^1$-hard, \cite{ad}, can be reduced to these problems. 
\nl  For more information about the analytical hierarchy 
(containing in particular the class $\Pi_1^1$) see  the textbook \cite{rog}.

\hs We now consider first the problem of determinizability or 
regular  complementability for timed regular $\om$-languages. 

\begin{theorem}\label{TBA}
The following problems are $\Pi_1^1$-hard. 
\nl  For a given TBA $\mathcal{A}$, determine whether :
\begin{enumerate}
\ite $L_\om(\mathcal{A})$ is accepted by a deterministic TBA.
\ite $L_\om(\mathcal{A})^c$ is accepted by a TBA. 
\end{enumerate}
\end{theorem}

\proo  Let $\Si$ be a finite alphabet and let $a\in \Si$. 
Let, as in Section 3,  $A$ be the set of timed words 
 containing only letters $a$ and such that there is a pair of  $a$'s which are separated 
by a time distance $1$. The timed language $A$ is regular but its  complement 
is not   timed  regular  \cite{ad}. 

\hs We shall use the $\Pi_1^1$-hardness of the universality problem for timed regular $\om$-languages: 

\hs Let $c$ be an additional  letter not in $\Si$. 
 For a given timed regular $\om$-language $L \subseteq (\mathcal{R} \times \Si)^\om$, 
we can construct another 
timed language $\mathcal{L}$ over the alphabet $\Ga=\Si\cup\{c\}$ defined as the union of 
the following three languages. 

\begin{itemize}
\ite $\mathcal{L}_1 = A.(\mathcal{R} \times \{c\}).(\mathcal{R} \times \Si)^\om$, 
where $A$ is the above defined timed regular 
language over the alphabet $\Si$. 
\ite  $\mathcal{L}_2$ is the set of infinite timed words over $\Ga$ having  no $c$'s or having at 
least two $c$'s. 
\ite $\mathcal{L}_3 = (\mathcal{R} \times \Si)^\star.(\mathcal{R} \times \{c\}).L$. 
\end{itemize}  

\hs The timed $\om$-language  $\mathcal{L}$ is regular because $L$ is a regular timed  $\om$-language 
and $A$ is a  regular timed language. 
 There are  now two cases. 
\begin{enumerate}

\ite[(1)] {\bf First case.}  $L = (\mathcal{R} \times \Si)^\om$. 
Then $\mathcal{L}= (\mathcal{R} \times (\Si\cup\{c\}))^\om$. 
Therefore $\mathcal{L}$ has the minimum possible complexity and it 
 is  accepted by a deterministic TBA (without any clock). 
Moreover its complement $\mathcal{L}^c$ is empty thus it is also accepted by a deterministic  TBA
(without any clock). 

\ite[(2)] {\bf Second case.} $L$ is strictly included into  $(\mathcal{R} \times \Si)^\om$, i.e. $L^c$ is non-empty. 
It is then easy to see that : $$ \mathcal{L}^c = A^c.(\mathcal{R} \times \{c\}).L^c$$
\noi where $ \mathcal{L}^c = (\mathcal{R} \times \Ga)^\om - \mathcal{L}$, 
$A^c=(\mathcal{R} \times \Si)^\star - A$, and  $L^c= (\mathcal{R} \times \Si)^\om - L$.

\hs   We are going to show that $\mathcal{L}^c$ is not timed $\om$-regular. Assume on the contrary that there is a TBA  
$\mathcal{A}$ such that $\mathcal{L}^c=L_\om(\mathcal{A})$.  
Consider the reading of a timed $\om$-word of the form $x.1.c.u$, where $x \in (\mathcal{R} \times \Si)^\star$  and $u \in (\mathcal{R} \times \Si)^\om$, 
by the TBA $\mathcal{A}$. When reading the initial segment $x.1.c$, the TBA $\mathcal{A}$ has to check that $x\in A^c$, i.e. that 
no pair of  $a$'s  in $x$ is separated by a time distance $1$; this is clearly not  
 possible for the same reasons which imply that $A^c$ is not timed regular (see above Section 3). 
Thus  $\mathcal{L}^c$ is not timed $\om$-regular. This implies that  $\mathcal{L}$ is not accepted by any deterministic TBA because the 
class of deterministic regular timed $\om$-languages is closed under complement, \cite{ad}. 
\end{enumerate}

\noi In the first case  $\mathcal{L}$ is accepted by a deterministic TBA and $\mathcal{L}^c$ is timed $\om$-regular. 
In the second case $\mathcal{L}$ is not accepted by any deterministic TBA and $\mathcal{L}^c$ is not  timed $\om$-regular. 
\nl This ends the proof because  the universality 
problem for timed B\"uchi automata is $\Pi_1^1$-hard, \cite{ad}. 
\ep 

\hs As in the case of TA reading finite length timed words, we can consider the corresponding problems with 
``bounded resources".

\hs  Below $TBA(n, K) $ denotes the class  of timed B\"uchi automata having 
at most $n$ clocks, where constants are at most $K$.

\hs Problem A.  Given a TBA $\mathcal{A}$ and non-negative integers $n, K$, does there exist a TBA
$\mathcal{B} \in  TBA(n, K)$ such that $L_\om(\mathcal{B})^c = L_\om(\mathcal{A})$ ? 

\hs Problem B.   Given a TBA $\mathcal{A}$ and non-negative integers $n, K$, does there exist a
deterministic TBA $\mathcal{B} \in TBA(n, K)$  such that $L_\om(\mathcal{B}) = L_\om(\mathcal{A})$ ? 

\hs 
We can infer from the proof of preceding Theorem,  that these problems are also $\Pi_1^1$-hard, 
because we have seen that, in the first case, $\mathcal{L}$ and $\mathcal{L}^c$ are accepted by deterministic 
timed B\"uchi automata {\it without any clock}.  

\hs In a very similar manner, using the same ideas as in the proof of Theorems 
\ref{min} and  \ref{TBA}, we can study the problem of minimization of the number of clocks 
for timed B\"uchi automata. We can then  show that it is $\Pi_1^1$-hard,  by reducing  to it the universality 
problem for timed B\"uchi automata with $n$ clocks, where $n\geq 2$, which  is $\Pi_1^1$-hard.  So we get the following result.

\begin{theorem} Let $n \geq 2$ be a positive integer. 
It is  $\Pi_1^1$-hard to determine, for a given TBA $\mathcal{A}$ with $n$ clocks, whether  there exists a TBA $\mathcal{B}$ with $n-1$
clocks, such that $L_\om(\mathcal{B}) = L_\om(\mathcal{A})$.
\end{theorem}

\begin{Rem}
We have already mentioned that, for timed automata with only one clock,  the universality problem is decidable   \cite{ow}.  
On the other hand, for timed B\"uchi automata with only one clock,  the universality problem 
has been recently shown to be   undecidable by  P. A. Abdulla, J. Deneux, J. Ouaknine,  and  J. Worrell in \cite{adow}. 
However it seems to us that,  in the paper  \cite{adow}, this problem is just proved to be undecidable and not $\Pi_1^1$-hard. 
Then we can just infer that the above theorem is still true for $n=1$ if we replace ``$\Pi_1^1$-hard" by ``undecidable". 
\end{Rem}

\hs {\bf Acknowledgements.}
Thanks  to the anonymous referees for useful comments
on a preliminary version of this paper.


\begin{thebibliography}{99}

\bibitem[AD94]{ad} R. Alur and D. Dill, 
A Theory of Timed Automata, Theoretical Computer Science, Volume 126, 
p. 183-235, 1994.


\bibitem[AM04]{am}
R. Alur and P. Madhusudan, 
Decision Problems for Timed Automata: A Survey, in 
Formal Methods for the Design of Real-Time Systems, International School on Formal Methods for the Design of Computer, 
Communication and Software Systems, SFM-RT 2004, Revised Lectures. 
Lecture Notes in Computer Science, Volume 3185,  Springer, 2004,  p. 1-24.

\bibitem[ADOW05]{adow}P. A. Abdulla, J. Deneux, J. Ouaknine,  and  J. Worrell, 
Decidability and Complexity Results for Timed Automata via Channel Machines,  in the Proceedings of the 
 International Conference  ICALP 2005, Lecture Notes in Computer Science, Volume  3580,  Springer,  2005, p. 1089-1101. 

\bibitem[Asa04]{asa} E. Asarin, Challenges in Timed Languages, From Applied Theory to 
Basic Theory, Bulletin of the European Association for Theoretical Computer Science, Volume 
83, p. 106-120, June 2004. 

\bibitem[ACM02]{acm} E. Asarin, P. Caspi, and O. Maler,  Timed Regular Expressions, Journal of the ACM, Volume 49 (2), 2002, p. 172-206. 


\bibitem[Dim05]{dim} C. Dima, 
Timed Shuffle Expressions,  
in the Proceedings of the  16th International Conference on  Concurrency Theory, CONCUR 2005, 
Lecture Notes in Computer Science, Volume 3653,  Springer,  2005,
p. 95-109. 


\bibitem[Fin03b]{rel-dec} O. Finkel, 
Undecidability of Topological and Arithmetical Properties
of Infinitary Rational Relations, 
RAIRO-Theoretical Informatics and Applications, Volume 37 (2), 2003,  p. 115-126.

\bibitem[Fin05]{beatcs1} O. Finkel, 
On Decision Problems for Timed Automata, 
Bulletin of the European Association for Theoretical Computer Science, Volume 87,  2005,  p. 185-190. 

\bibitem[Fin06]{beatcs2} O. Finkel,  On the Shuffle of Timed Regular languages, 
Bulletin of the European Association for Theoretical Computer Science, Volume 88,  2006,  p. 182-184.

\bibitem[HKW95]{hkw} T. A. Henzinger, P. W. Kopke, and  H. Wong-Toi,  The Expressive Power of Clocks, 
in the Proceedings of the 22nd International Colloquium, ICALP95,  Lecture Notes in Computer Science, Volume 944,  Springer,  1995, 
p. 417-428. 


\bibitem[OW04]{ow} J. Ouaknine and J. Worrell, 
 On the Language Inclusion Problem for Timed Automata: Closing a Decidability Gap, in the Proceedings of the 
19th Annual IEEE Symposium on Logic in Computer Science, LICS  2004, 
Turku, Finland,  IEEE Computer Society,   2004, p. 54-63. 

\bibitem[Rog67]{rog} H. Rogers, Theory of Recursive Functions and Effective Computability, 
McGraw-Hill, New York, 1967.

\bibitem[Tri04]{tri} S. Tripakis, Folk Theorems on the Determinization and 
Minimization of Timed Automata, in the Proceedings of FORMATS'2003, Lecture Notes in Computer 
Science, Volume 2791, p. 182-188, 2004. 



\end{thebibliography}
\end{document}